\documentclass[usenatbib]{mn2e}
\usepackage{graphicx,fixltx2e}
\usepackage{wasysym}
\usepackage{color}

\def\kms{\,km\,s$^{-1}$}         
\def\mjup{M$_{Jup}$}             

\def\rpd{\hbox{rad\,d$^{-1}$}}

\def\chisqr{\hbox{$\chi^2_{\rm r}$}}
\def\mSun{\hbox{${\rm M}_{\odot}$}}

\def\rSun{\hbox{${\rm R}_{\odot}$}}

\def\mstar{\hbox{$M_{\star}$}}
\def\rstar{\hbox{$R_{\star}$}}
\def\teff{\hbox{$T_{\rm eff}$}}
\def\logg{\hbox{$\log g$}}

\def\ms{\hbox{m\,s$^{-1}$}}

\def\kms{\hbox{km\,s$^{-1}$}}
\def\vsini{\hbox{$v \sin i$}}

\def\degr{\hbox{$^\circ$}}

\def\omeq{\hbox{$\Omega_{\rm eq}$}}
\def\dom{\hbox{$d\Omega$}}

\begin{document}

\title[Magnetic fields of planet-host stars]
{A small survey of the magnetic fields of planet-host stars\thanks{Based on observations obtained by NARVAL at T\'elescope Bernard Lyot (CNRS) and ESPaDOnS at Canada-France-Hawai telescope.}}

\makeatletter

\def\newauthor{%
  \end{author@tabular}\par
  \begin{author@tabular}[t]{@{}l@{}}}
\makeatother

\author[R.~Fares et al.]{\vspace{1.7mm} 
R.~Fares$^{1}$\thanks{E-mail:
rf60@st-andrews.ac.uk},
 ,C.~Moutou$^2$, J.-F.~Donati$^3$, C.~Catala$^4$, E. Shkolnik$^5$, M.M.~Jardine$^1$ \\ 
\vspace{1.7mm}
{\hspace{-1.5mm}\LARGE\rm
 A.C.~Cameron$^1$, M.~Deleuil$^2$}\\  
$^1$ School of Physics and Astronomy, Univ.\ of St~Andrews, St~Andrews, Scotland KY16 9SS, UK \\
$^2$ Aix Marseille Universit\'e , CNRS, LAM, UMR 7326, F-13388, Marseille, France\\
$^3$ IRAP-UMR 5277, CNRS \& Univ. de Toulouse, 14 Av. E. Belin F-31400 Toulouse, France \\
$^4$ Observatoire de Paris, 61 avenue de l'Observatoire F-75014 Paris, France\\
$^5$ Lowell Observatory, 1400 West Mars Hill Road, Flagstaff, AZ 86001, USA \\ 
}

\date{2013, MNRAS, submitted}
\maketitle

\begin{abstract}
Using spectropolarimetry, we investigate the large-scale magnetic topologies of stars hosting close-in exoplanets. A small survey of ten stars has been done with the twin instruments TBL/NARVAL and CFHT/ESPaDOnS between 2006 and 2011. Each target consists of circular-polarization observations covering 7 to 22 days. For each of the 7 targets in which a magnetic field was detected, we reconstructed the magnetic field topology using Zeeman-Doppler imaging. Otherwise, a detection limit has been estimated. Three new epochs of observations of $\tau$~Boo are presented, which confirm magnetic polarity reversal. We estimate that the cycle period is 2 years, but recall that a shorter period of 240 days can not still be ruled out. The result of our survey is compared to the global picture of stellar magnetic field properties in the mass-rotation diagram. The comparison shows that these giant planet-host stars tend to have similar magnetic field topologies to stars without detected hot-Jupiters. This needs to be confirmed with a larger sample of stars. 
\end{abstract}


\section{Introduction}

The role of the stellar magnetic field in the evolution of stellar and planetary systems is suspected to be important, but poorly constrained by observations. For instance, stellar magnetic braking, planet migration, and dynamical evolution may be acting simultaneously in the early stages of evolution of the systems, with an impact on the final state that depends on the system properties \citep{dobbs04, lai11}. In the case of short-period planets, the interactions between the planet and the star continue throughout the lifetime of the system, as the planet may be embedded in the magnetosphere of the star at only a few stellar radii from the star's surface. The impact of the stellar wind may then be important \citep{vidotto}, and even reconnections between the stellar and planetary magnetic fields could happen \citep{cohen10, cohen11, lanza12}. This can influence the planetary magnetic field, the planetary upper atmosphere and maybe the internal structure of the planet as well. On the stellar side, the close-in planet, especially when it is massive, may induce anomalies on the stellar surface through magnetic interactions (\citealt{shk03,shk05,walker08,pagano09}), although several observational searches for such signatures give results that are either unconfirmed, intermittent or difficult to interpret (e.g., \citet{cranmer07, shk08, fares12}). In order to better understand the environment where a close-in planet orbits its star, it is necessary to have information about the stellar magnetic field topology. Then, extrapolation techniques may be used to get quantified properties of the magnetic environment that may impact the planet \citep{jardine02,fares10,fares12}. \\

In this paper, we investigate the large-scale magnetic properties of ten planet-host stars using spectropolarimetric observations, in order to provide inputs to 1) more intensive similar campaigns on stars where the magnetic field is strong enough for an accurate characterization, and 2) extrapolation models that explore the star-planet magnetic interactions. Three new epochs of observations are reported for the short-cycle $\tau$~Boo star, which makes 8 the total number of epochs when we observed this star. In Section \ref{secobs}, we describe the observational method and material, in Section \ref{secstars} we present the stellar sample, and in Section \ref{secdiscu} we discuss the results before concluding.

\section{Observations}
\label{secobs}
We have secured spectropolarimetric observations of stars hosting close-in extrasolar planets, using either ESPaDOnS at the 3.6-m Canada-France-Hawaii Telescope on Mauna Kea or NARVAL at the 2-m Telescope Bernard Lyot in Pic du Midi (France). Both instruments are twin high-resolution spectropolarimeters that measure the circular polarisation in stellar spectral lines using multiple exposures. The spectral resolution and range are respectively $65000$ and 370-1000 nm in the polarisation mode. Four exposures per observation are necessary to derive the circular polarisation (Stokes V) profiles and check its significance with respect to spurious polarisation signals. The data were collected between June 2006 and January 2011, over a sample of 10 planet-host stars. 
For some stars of our sample, the number of collected spectra is limited, as they correspond to a first-investigation survey for spectropolarimetric detections in preparation for more intensive follow-up observations of detected fields. Table \ref{obslog} gives a summary of the observations performed in this program.

\begin{table*}
\vspace{1cm}
\caption{Summary of observations and results presented in this paper. The instrument used is either CFHT/ESPaDOnS (ESP) or TBL/NARVAL (NAR). The epochs of observations and numbers of spectra ($\#$ seq) are listed. Detection status is N: no detection, Y: few detections, M: several detections and map reconstruction. The mean magnetic field strength (B), the percentage of poloidal energy over total magnetic energy (Pol) and the percentage of axisymmetry in the poloidal field are given for all epochs. When there is no detection, an upper limit is given for B for an adopted peculiar magnetic topology (and thus poloidal contribution).}          
\begin{tabular} {lcccccccc} 
\hline
Name & Instrument & Date & $\#$ seq & Detection &B (G) & \% Pol&\% Axisy  & Ref\\ 
\hline
HD 46375    & ESP  & Jan08 &10&Y-M& 2&99&78& this work\\
HD 46375    & NAR & Sep08 & 18 & M &3.2 &&& \cite{gaulme10}\\
HD 73256   & ESP  & Jan08 &9&M&2.7 &80&4& this work\\
HD 102195 & ESP  & Jan08 &10&M& 12.4&44&25&  this work\\
HD 130322 & ESP & Jan08 &9&M&2.5 &84&58& this work\\
HAT-P-2 & ESP & Jan08 &4&N& $<$40&75&-& this work\\
HD 179949   & ESP & Jun07&19&M&2.6 &80&55& \cite{fares12}\\
HD 179949   & ESP & Sep09&10&M&3.7 &90&36&\cite{fares12}\\
HD 189733   & ESP & Jun06 & 19 & M &33  &77&56& \cite{moutou07}\\
HD 189733   & NAR & Jun07 & 20 &M &22 &43&26& \cite{fares10}\\
HD 189733   & NAR & Jul08 & 24&M &36 &33&17& \cite{fares10}\\
CoRoT-7     & NAR & Jan10 &4& N & $<$150 &100&-& this work\\
$\tau$ Bootis& ESP & Jun06 & 12&M& 1.8&&& \cite{catala07}\\ 
$\tau$ Bootis& NAR/ESP & Jun07 & 2/30&M&3.7 &83&60& \cite{donati08}\\ 
$\tau$ Bootis& ESP & Jan08 & 40&M&3.1 &38&20& \cite{fares09}\\ 
$\tau$ Bootis& NAR & Jun08 & 9&M& 2.3&87&36& \cite{fares09}\\ 
$\tau$ Bootis& NAR & Jul08 & 19&M& 1.7&91&62& \cite{fares09}\\ 
$\tau$ Bootis& NAR & Jun09 &&M&2.7 &88&43& this work\\ 
$\tau$ Bootis& NAR & Jan10 &&M&3.8 &62&46& this work\\ 
$\tau$ Bootis& NAR & Jan11 &&M&3.2 &70&37& this work\\ 
XO-3           & ESP & Jan08 &15&N&$<$20 &90&-& this work\\
\hline
\end{tabular}
\label{obslog}  
\end{table*}

The data were reduced with the software {\sc libre-esprit} that automatically extracts and calibrates intensity and polarization spectra. The Least-Square-Deconvolution (LSD, \citealt{donati97}) profiles are calculated to significantly improve the SNR, using a mask adapted to the spectral type of each target. On average, more than 6000 stellar lines are used to produce these intensity and polarisation profiles. The LSD profiles are corrected for the radial-velocity shift of the star, including the motion due to the planet. The radial-velocity precision of the stellar intensity profiles are better than 30 \ms.

\section{Stellar properties}
\label{secstars}
The stellar sample selected for our study includes 10 stars brighter than $V=12$, hosting planets at orbital periods less than 11 days. Most of these planets are giant planets with masses larger than 0.22~M$_{Jup}$, except CoRoT-7 b which is a telluric planet in an extremely short orbit (0.015~M$_{Jup}$ and 0.85 day period, \citealt{leger09}). The stellar parameters adopted in this work are summarized in Table \ref{stars}. The rotation periods are a critical parameter, and often the least constrained one. The targets were originally selected for their short rotational periods, in order to allow observations with the two-week runs with ESPaDOnS; this does not apply, however, for HD 46375 and CoRoT-7, which were selected because of existing data of the CoRoT satellite (despite their long rotational periods). Note also that the rotation period of HD 130322 was recently updated to 26 days \citep{simpson10} while it was given as 12 days in the planet discovery paper \citep{udry00}. There is also contradiction in the literature about the rotation period of HD 102195 (12 days in \citet{ge06} and 20 days in \citet{melo07}): in our analysis, we choose the 12-day value which is based on photometric observations rather than on activity calibrations.

Concerning the stellar inclination, we use in general the value derived from the chosen rotation period, with $\sin i= \vsini \times P_{rot} / (2 \pi \rstar)$. The reconstruction of the stellar magnetic field is, however not very sensitive to a precise knowledge of the inclination (up to 20\degr, \citealt{morin08}).

\begin{table*}
\vspace{1cm}
\caption{Fundamental parameters of stars used in this work and some properties of their planets. The columns list the name of the star, its visual magnitude, spectral type, effective temperature, log of the gravity at the surface, [Fe/H], stellar mass, stellar radius, \vsini, rotation period, orbital period of the planet, semi-major axis of the planetary orbit, planet's projected mass and the references.}
\begin{tabular} {lccccccccccccc} 
\hline
Name & Vmag & SpT & \teff & \logg &[Fe/H]& \mstar &\rstar & \vsini & P$_{\rm rot}$ & P$_{\rm orb}$ &$a$& M$_p \sin i$  &Ref.\footnotemark[1] \\
	&	&		&K&		&	   &\mSun&\rSun&\kms&days&days&AU&\mjup&\\
\hline
HD 46375 & 7.9&K1IV & 5290 & 4.66 & 0.39& 0.97 & 0.86 & 1.2 & 42 & 3.0236&0.0399&0.2272&G10,B06\\
HD 73256 & 8.08 & G8 & 5636 & 4.30& 0.26& 1.05 & 0.89 & 3.2& 14 &2.5486&0.0371&1.869&B06,U03\\
HD 102195 & 8.05 & K0V & 5290 & 4.45 & 0.05 & 0.87 & 0.82 & 2.9 & 12.3  &4.1138&0.0479&0.453& G06,M07\\
HD 130322 & 8.04 & K0V & 5330 & 4.41& -0.02 & 0.79 & 0.83 &1.6 & 26.1&10.708&0.0896&1.043&U00,S10 \\
HAT-P-2 & 8.71& F8 & 6290 & 4.22 & 0.12 & 1.36 & 1.64 & 20.8& 3.8&5.6335&0.0687&9.09&P10\\
HD 179949 &6.25&F8V&6168&4.34&0.14&1.21&1.19&7.0&7.6&3.0925&0.0439&0.902&B06,F12\\
HD 189733 &7.7&K2V&5050&4.59&-0.03&0.82&0.76&2.97&12.5&2.2186&0.0310&1.140&B05,F10\\
CoRoT-7 & 11.7 & G9V & 5250 & 4.47& 0.12 & 0.91 & 0.82 & 1.1 & 23.6 & 0.8536&0.0172&0.0151&  B10 \\
$\tau$ Bootis& 4.5 & F7V & 6387 & 4.25&0.23 & 1.34 & 1.42 & 15.0 & 3.0 &3.3124&0.0480&4.170& F09,B12\\
XO-3 & 9.8 & F5V & 6430 & 3.95& -0.18 & 1.41 & 2.08 & 18.3 & 3.7 &3.1915&0.0454&11.79&JK08,W09\\
\hline
\end{tabular}
{$^1$}G10: \cite{gaulme10},B06: \cite{butler06},U03: \cite{udry03},G06: \citet{ge06},M07: \citet{melo07}, U00: \citet{udry00},S10: \citet{simpson10},P10: \citet{pal10},F12: \citet{fares12},B05: \citet{bouchy05},F10: \citet{fares10},B10: \citet{bruntt},F09: \citet{fares09},B12: \citet{brogi},JK08: \citet{jk08},W09: \citet{winn09}
\label{stars}  
\end{table*}

\section{Spectropolarimetric analyses}

When a sufficient number of detected Stokes V profiles is available, we reconstruct the best-fit magnetic topology using a tomographic technique called Zeeman Doppler Imaging (ZDI) as developed by \citet{donati97, donati06} and described in these papers. ZDI consists of inverting series of Stokes V profiles into the stellar magnetic field topology responsible for producing these profiles. The problem is ill-posed, ZDI uses the principles of maximum entropy to retrieve the simplest image compatible with the data. The magnetic field is described by its radial, azimuthal and meridional components, all expressed in terms of spherical harmonics expansions. 
This description of the field allows to calculate easily the contribution of each spherical harmonic order to the field, as well as the contribution of the poloidal and toroidal components and the degree of axisymmetry. Note that the axisymmetric contribution is given by modes with $m<l/2$.

\label{secdiscu}
\subsection{The evolution of magnetic topology $\tau$~Boo} 
Previous spectropolarimetric analyses of 5 epochs of observations of $\tau$~Boo (F7V) have been described in \citet{catala07}, \citet{donati08} and \citet{fares09}. These data showed two occurrences of polarity reversals. The derived length of the magnetic cycle was about two years; the temporal sampling of the observations, however, allowed other possible values for the cycle period.

New data were acquired with TBL/NARVAL during three separate epochs, in May 2009 (data spanning 19 days), January 2010 (20 days) and January 2011 (12 days). The journal of these new observations is shown in appendix \ref{appendix} (Table \ref{obsjournaltb}).

The LSD profiles were calculated using the same method and parameters as in \citet{fares09}. We reconstructed the magnetic topology of $\tau$~Boo for all three epochs using the differential rotation (hereafter DR) measured by \citet{fares09}. We used up to 8 degrees of spherical harmonics and an inclination of 45$^{\circ}$. Our measurements of the inclination using ZDI in \citet{catala07}, \citet{donati08} and \citet{fares09} agrees with the measurments of \citet{rodler} and \citet{brogi} who used a different technique. We fit the $V$ profiles to a level of reduced chi-square $\chisqr=0.95$. The observed and fitted profiles are shown in  Figure \ref{fig:profiltb}. The maps projected on a spherical coordinate system are shown in Figure \ref{fig:maptb}.

\begin{figure*}
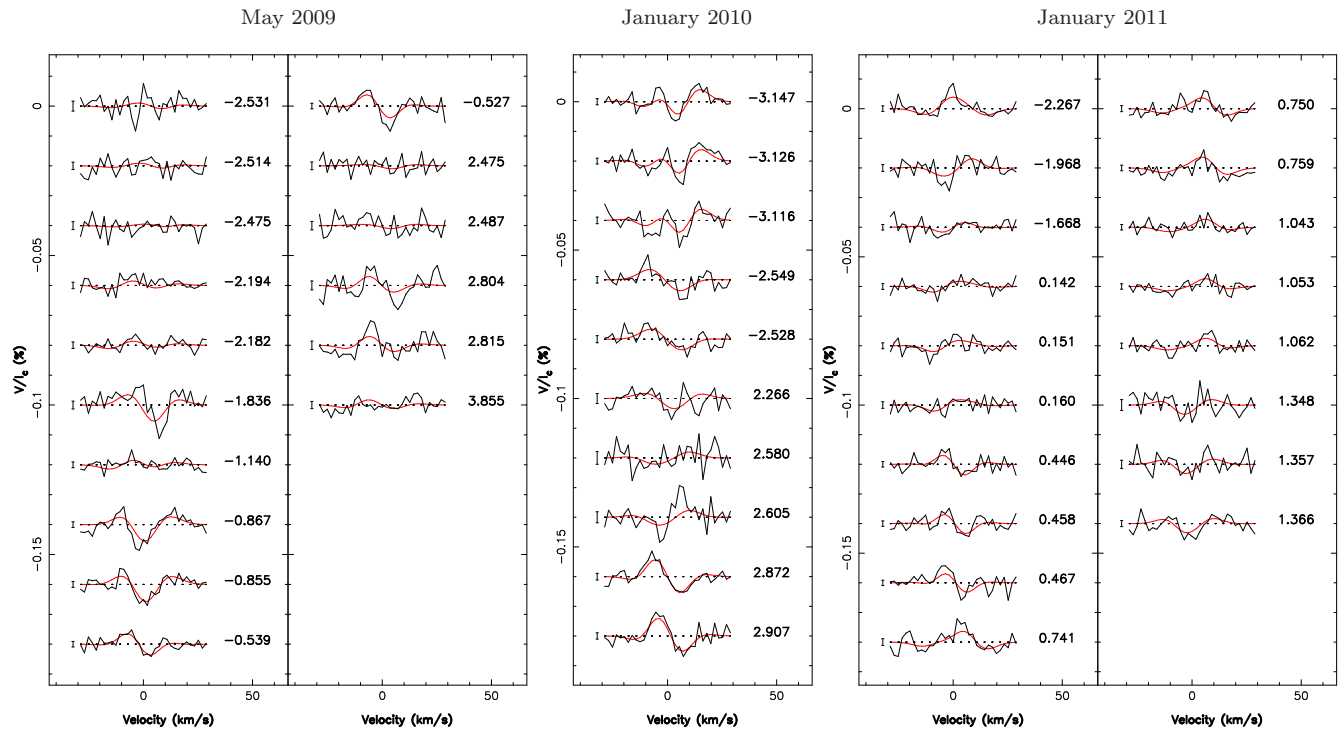

 \begin{minipage}[c]{1.\linewidth}
~~~~~~~~~~~~~~~~~~~~~~~~~~~~~~~May~2009~~~~~~~~~~~~~~~~~~~~~~~~~~~~~~~~~~~~~~January~2010~~~~~~~~~~~~~~~~~~~~~~~~~~~~~~~~~~~~~~January~2011\\
   \end{minipage} \hfill
\begin{minipage}[c]{1.\linewidth}
\center{\includegraphics[width=9cm,angle=270]{ProfilV_tb_may09.ps}
\includegraphics[width=9cm,angle=270]{ProfilV_tb_jan10.ps}
\includegraphics[width=9cm,angle=270]{ProfilV_tb_jan11.ps}}
\end{minipage} \hfill
\caption{Circular polarization profiles of $\tau$~Boo obtained in May 2009, January 2010 and January 2011 with TBL/NARVAL. The observed and synthetic 
profiles are shown in black and red, respectively. On the left of each profile we show a $\pm$1$\sigma$ error bar, while on the right the rotational cycles are indicated. }
\label{fig:profiltb}
\end{figure*}

\begin{figure*}
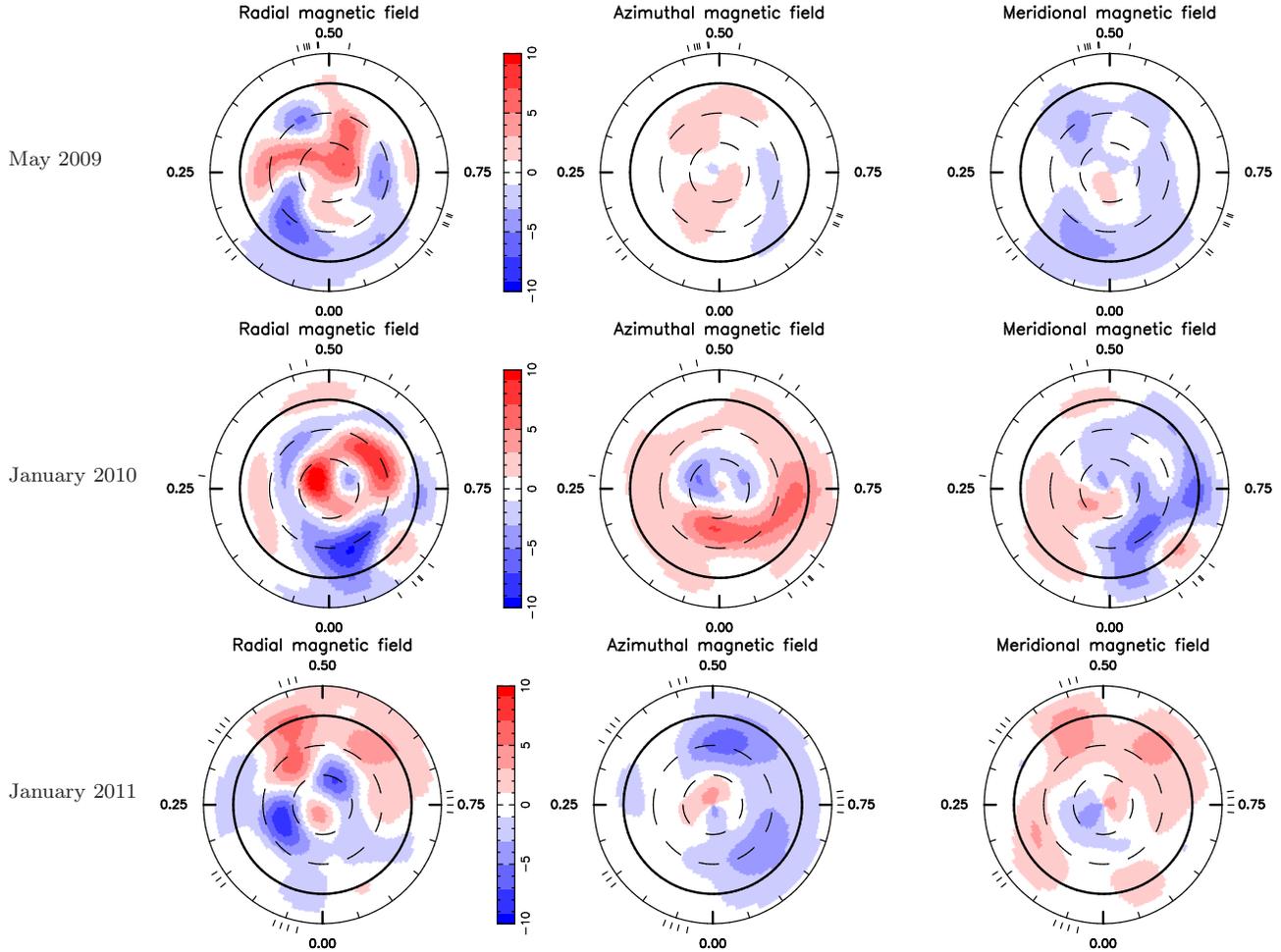

\begin{minipage}[c]{0.11\linewidth}
May~2009
\end{minipage} \hfill
 \begin{minipage}[c]{0.88\linewidth}
\includegraphics[width=15cm]{map_may09_tauboo.ps}
\end{minipage} \hfill
\begin{minipage}[c]{0.11\linewidth}
January~2010
\end{minipage} \hfill
 \begin{minipage}[c]{0.88\linewidth}
\includegraphics[width=15cm]{map_jan10_tauboo.ps}
\end{minipage} \hfill
\begin{minipage}[c]{0.11\linewidth}
January~2011
\end{minipage} \hfill
 \begin{minipage}[c]{0.88\linewidth}
\includegraphics[width=15cm]{map_jan11_tauboo.ps}
\end{minipage} \hfill
\caption{Magnetic topology of $\tau$ Boo reconstructed from profiles in Figure \ref{fig:profiltb}: May 2009 (top row); January 2010 (middle row); January 2011 (bottom row).  The radial, azimuthal and meridional components of the field (with magnetic field strength labelled in G) are depicted. The star is shown in flattened polar projection down to latitudes of 30\degr, with the equator depicted as a bold circle and parallels as dashed circles. Radial ticks around each plot indicate rotational phases of observations. }

\label{fig:maptb}
\end{figure*}

Previous observations have shown a large DR in $\tau$ Boo, with a surface shear of the order of $\dom=0.4\pm0.1$~\rpd\ and an equatorial angular velocity of $\omeq=2.0\pm0.1$~\rpd\ \citep{donati08,fares09}. DR is again detected in the May 2009 data set, with corresponding values of $\omeq=1.98\pm0.01$~\rpd and $\dom=0.15\pm0.03$~\rpd. The value of \dom\ is significantly smaller than values measured in previous epochs. Although our observations cover 20 days, the rotational phases do not sample the stellar surface very widely, which may induce a bias in deriving the DR. For this reason, we reconstrucetd the maps using the DR parameters as measured in previous epochs, for all data.

The properties of the reconstructed magnetic maps for the three new epochs are summarized in Table \ref{tab:tbmap}. The average magnetic field ranges from 2.7 to 3.8 G, with values very similar to the ones reported in earlier analyses (1.7 to 3.7 G in \citealt{fares09}). The contribution of the toroidal component to the total magnetic energy varies from 12 to 30\%, in a smaller extent with respect to earlier epochs (9 to 62\%). The last epoch of observation, in January 2011, shows a new polarity reversal compared to January 2010. In addition, the field has also switched polarity between July 2008 (last map in \citealt{fares09}) and May 2009.  As observed earlier, the field configuration evolves inside a cycle: between May 2009 and January 2010, the energy distributed in the radial field has decreased while the energy in the azimutal field has increased.

\subsection*{Period of magnetic cycle} 

In order to determine the length of $\tau$~Boo's magnetic cycle, we performed a period search similar to the one described in \cite{fares09}. We calculated the (signed) magnetic flux in both the radial and azimuthal component of the field for each reconstructed map. In the case of the radial field, the flux is counted positive for latitudes greater than 30\degr, and negative for latitudes between 0\degr and 30\degr. We then simultaneously fitted the two fluxes with two sine waves of equal period, the period being varied on a range of 100-1300 days. We found, as in \cite{fares09}, two periods that fit the data well. The first one is of 740 days (2 years), and the second one is of 240 days (8 months), see Figure \ref{fig:cycle}. We then calculated the false-alarm probability (FAP) of these periods. We produced 10000 data sets by night shuffling, and fitted each data set following the same procedure described above. The  FAP is the number of data sets for which the \chisqr~is smaller than the \chisqr~of our periods divided by 10000. We find a FAP of 3\% for the 240 day period, and a FAP of 15\% for the 740 day period. 

\begin{figure*}
\center{\hbox{\includegraphics[scale=0.4]{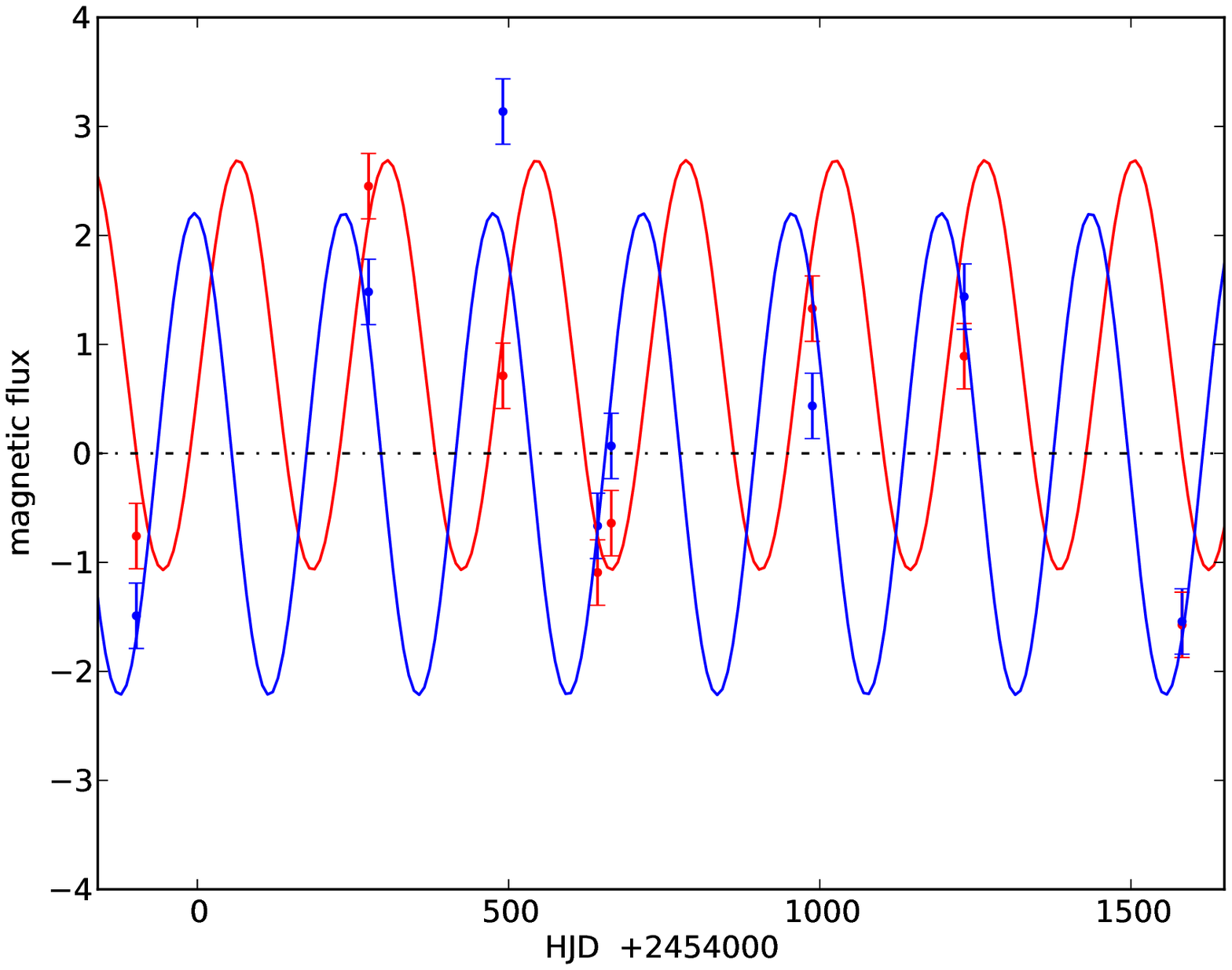}\hspace{3mm}\includegraphics[scale=0.4]{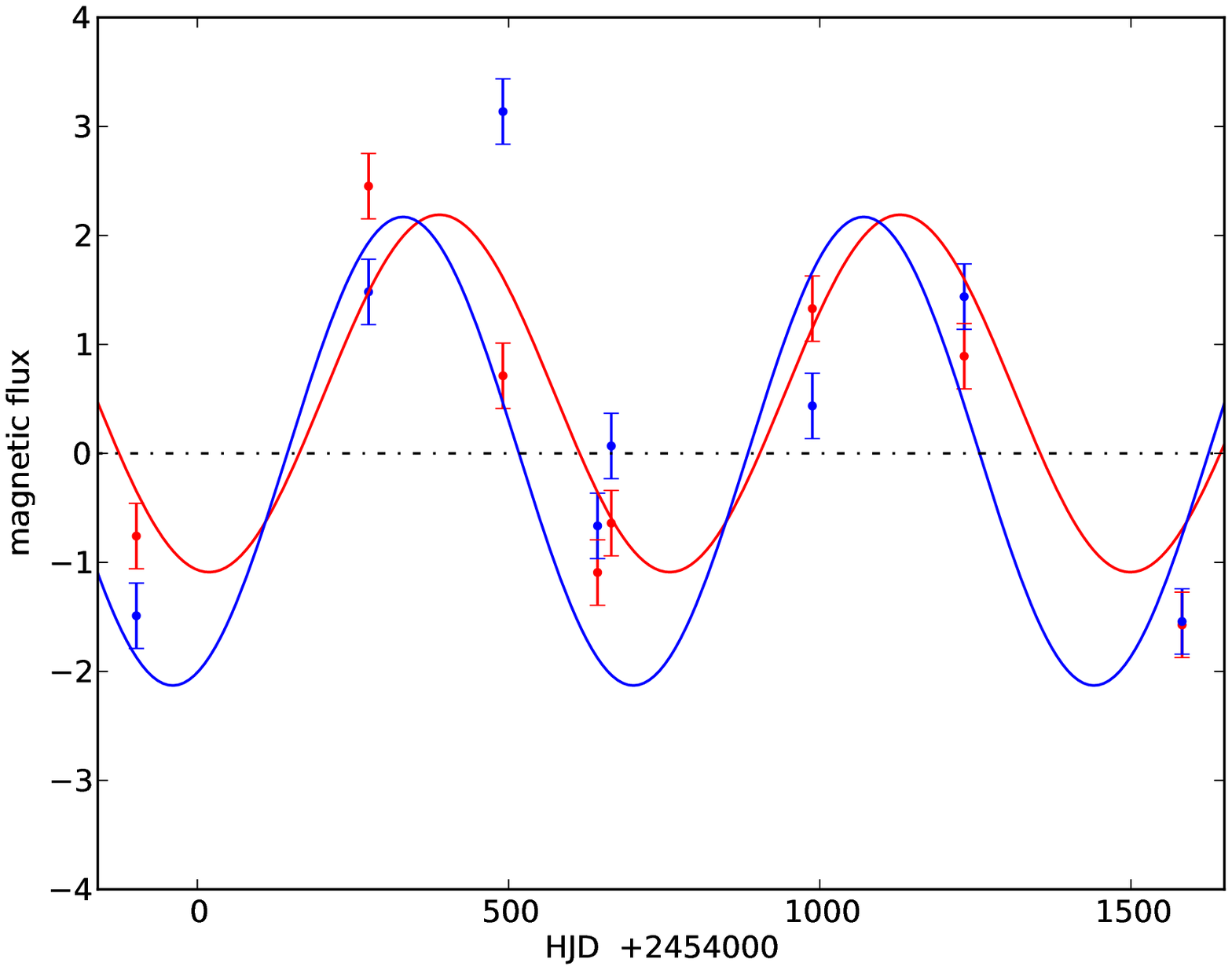}}}
\caption{Fluxes of the radial field (red) and azimuthal field (blue) vs HJD, calculated for the Northern hemisphere of $\tau$~Boo. In the particular case of $B_{r}$, the magnetic flux is counted positive for latitudes superior to 30\degr~and negative for latitudes between 0\degr~and 30\degr~to take into account the contribution of both dipolar and quadrupolar terms of the poloidal field (as in  \protect\citealt{fares09}). The best sinusoidal fit for $\rm P=240$~d (left pannel) and $\rm P=740$~d (right pannel) are plotted.}
\label{fig:cycle},
\end{figure*}

\begin{table*}
\caption{Summary of magnetic topology evolution of $\tau$ Boo: average magnetic field B, percentage of the toroidal energy relative to the total energy, percentage of the energy contained in the axisymmetric modes of the poloidal component (modes with $m<l/2$) and percentage of the energy contained in the modes of  $l \leq$2 of the poloidal component for each epoch of observations. } 
\begin{tabular} {lccccc} 
\hline
Epoch			&  B 	&\% toroidal & \% axisym in poloidal & \%  $l \leq$2 in poloidal & Reference\\ 
				& (G)	&	\%	  &	\%		&	\%		&\\
\hline
June 2007 & 3.7 & 17 & 60 & 52 & \cite{donati08}\\
January 2008 & 3.1 & 62 & 20 &50 &\cite{fares09} \\
June 2008 & 2.3 & 13 & 36& 36  &\cite{fares09} \\ 
July 2008 & 1.7 & 9 & 62& 47&\cite{fares09}  \\
May 2009			& 2.7		& 12			& 	43		&	47	&this work\\
January 2010		& 3.8		& 38			& 	46		&	40	&this work\\
January 2011		& 3.2		& 30			& 	37		&	50	&this work  \\
\hline
\end{tabular}
\label{tab:tbmap}  
\end{table*}

\subsection{HD 73256}
Nine ESPaDOnS spectra of HD 73256 (G8) spanning 11 days were obtained in January 2008. Six circular polarisation signatures are detected. 
The adopted stellar inclination is 75$^\circ$ deduced from the rotational period measured by photometry \citep{udry03}.
DR is not detected in the data and thus is fixed to zero for the reconstruction of the magnetic field.
When correcting each LSD profile for the radial velocity of the star, we found that our measurements did not match the orbital ephemeris published in the literature, $T_0$=2452500.18 $\pm$ 0.28 \citep{udry03}. We updated the orbit phase using a measured $T_0$=2452500.42. 
The magnetic map is reconstructed for a $\chisqr$ of 1.15, which produces a reasonable fit to the Stokes V profiles (Figure \ref{fig:profils} first column).
The magnetic field that best matches the observations is an 80\% poloidal field with mean strength of 2.7 G. A small fraction of the the poloidal field is in axisymmetric modes ($\sim$ 4\%). The reconstructed topology of the stellar surface field is shown on Figure \ref{fig:maps} (top row).

\begin{figure*}
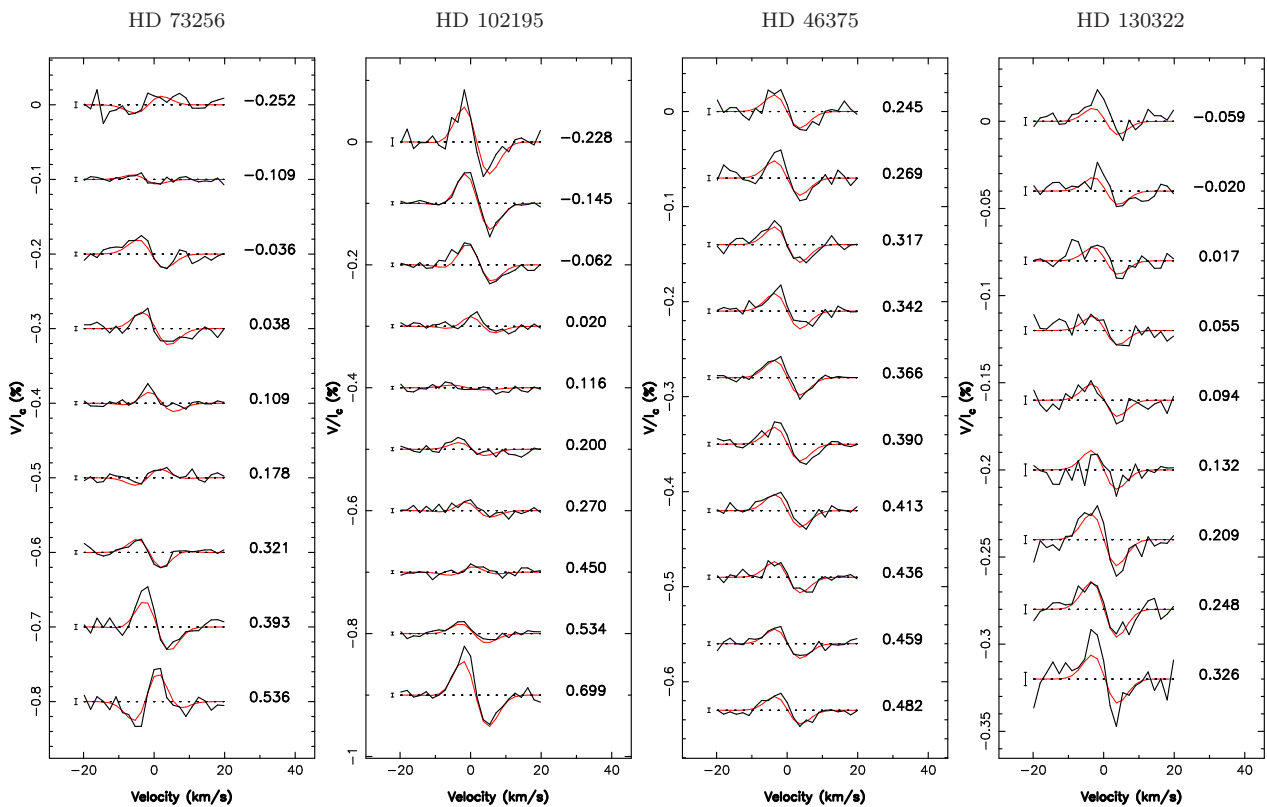

 \begin{minipage}[c]{1.\linewidth}
~~~~~~~~~~~~~~~~~~~~~HD~73256~~~~~~~~~~~~~~~~~~~~~~~~~~~~HD~102195~~~~~~~~~~~~~~~~~~~~~~~~~~~~~HD~46375~~~~~~~~~~~~~~~~~~~~~~~~~~~~~HD~130322\\

   \end{minipage} \hfill
\begin{minipage}[c]{1.\linewidth}
\center{\includegraphics[width=10cm,angle=270]{ProfilV_hd73256.ps}
\includegraphics[width=10cm,angle=270]{ProfilV_hd102195.ps}
\includegraphics[width=10cm,angle=270]{ProfilV_hd46375.ps}
\includegraphics[width=10cm,angle=270]{ProfilV_hd130322.ps}}
 \end{minipage} \hfill
\caption{Circular polarization profiles obtained in January 2008 with CFHT/ESPaDOnS for stars HD 73256, HD 102195, HD 46375 and HD 130322 respectively from left to right. See legend details in Figure \ref{fig:profiltb}.}
\label{fig:profils}
\end{figure*}

\begin{figure*}
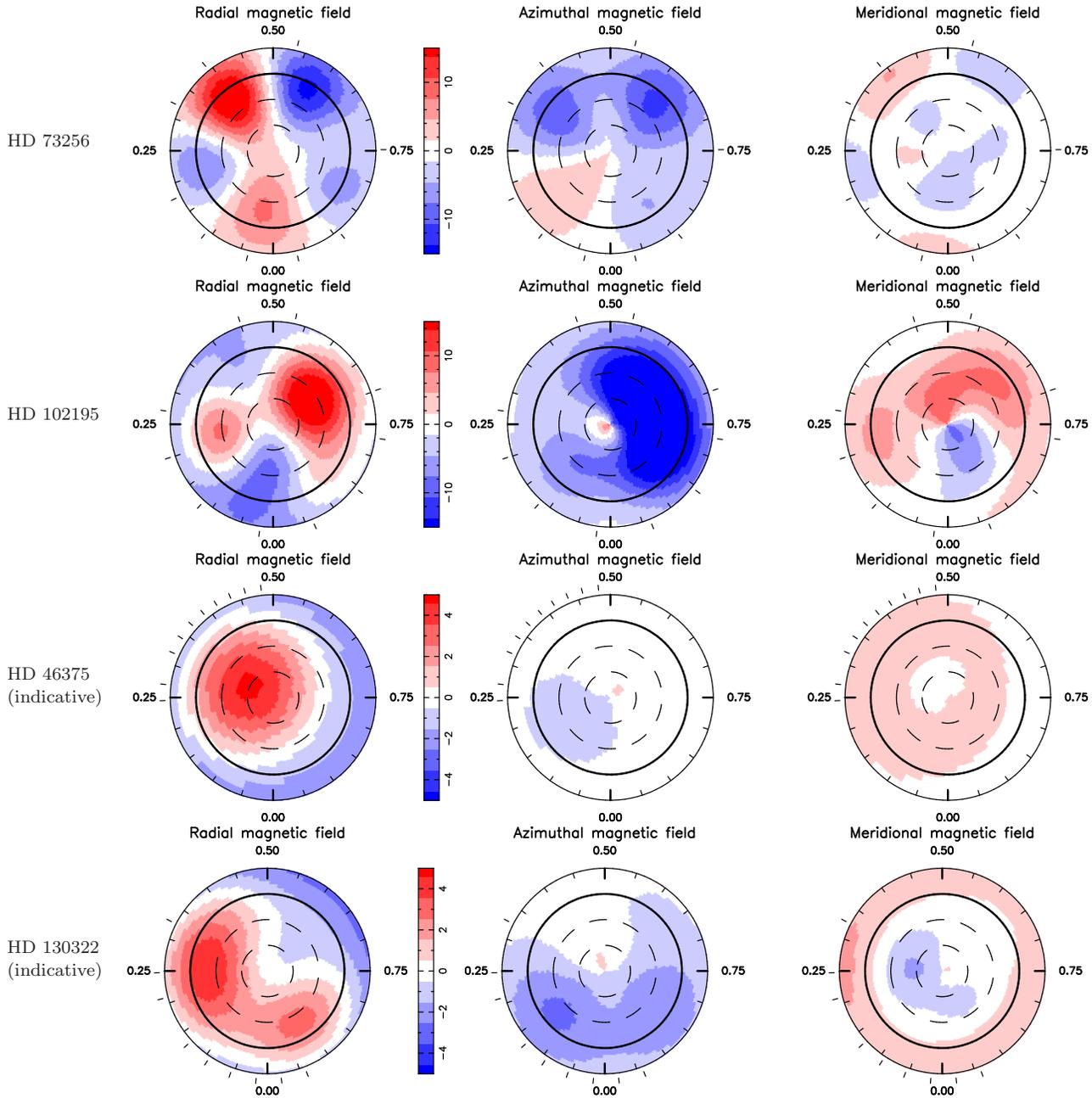

\begin{minipage}[c]{0.1\linewidth}
HD~73256
\end{minipage} \hfill
 \begin{minipage}[c]{0.89\linewidth}
\includegraphics[width=15cm]{map_hd73256.ps}
\end{minipage} \hfill
\begin{minipage}[c]{0.1\linewidth}
HD~102195
\end{minipage} \hfill
 \begin{minipage}[c]{0.89\linewidth}
\includegraphics[width=15cm]{map_hd102195.ps}
\end{minipage} \hfill
\begin{minipage}[c]{0.1\linewidth}
HD~46375\\
(indicative)
\end{minipage} \hfill
 \begin{minipage}[c]{0.89\linewidth}
\includegraphics[width=15cm]{map_hd46375.ps}
\end{minipage} \hfill
\begin{minipage}[c]{0.1\linewidth}
HD~130322\\
(indicative)
\end{minipage} \hfill
 \begin{minipage}[c]{0.89\linewidth}
\includegraphics[width=15cm]{map_hd130322.ps}
\end{minipage} \hfill
\caption{Magnetic maps of HD 73256, HD 102195, HD 46375 and HD 130322 from top to bottom rows. See legend details in Figure \ref{fig:maps}.}
\label{fig:maps}
\end{figure*}

\subsection{HD 102195}
\label{102195}
Ten spectra of HD 102195 (K0V) were obtained with ESPaDOnS in January 2008 (spanning 11 days), among which eight show a definite detection of the magnetic field.
The data do not show any evidence for DR, we consider a solid rotation at the 12 day period measured by \citet{ge06} and an inclination of 50\degr\ for the magnetic field reconstruction. The radial velocity of the star is corrected from the systemic velocity and the planet induced motion, as given in the literature \citep{melo07}. Our ESPaDOnS observations are not of sufficient velocity precision to permit the detection of the planet signal, especially on this star that exhibits activity jitter \citep{melo07}. 

Our observations cover almost one stellar rotation. The circular polarisation profiles are well fitted as shown in Figure \ref{fig:profils} (second column). The field modelling is achieved for a $\chisqr=1.2$. The characteristics of the best-fit magnetic model features a dipole contributing by 70\% to the poloidal component. The field's mean strength is of 12.5 G, 45\% of the magnetic energy in the poloidal field (Figure \ref{fig:maps}, second row), and 25\% of the poloidal field is axisymmetric.

\subsection{HD 46375} 
HD 46375 (K1IV) has been first observed with ESPaDOnS in January 2008, and then with NARVAL in October 2008. This later data set has been obtained simultaneously with CoRoT photometric observations and is described in \citet{gaulme10}. We describe in the following the data obtained with ESPaDOnS, although its temporal coverage is much poorer: only one quarter of the rotational period has been covered. Such a poor sampling of the stellar surface prevents the reconstruction of a magnetic map. We just recall the properties of the field as characterized by the NARVAL observations in October 2008: the field is dominated by a slightly tilted and mostly axisymmetric dipole with respect to the rotation axis; the magnetic strength at the pole is of the order of 5 G. The Stokes V profiles observed in January 2008 with ESPaDOnS (Figure \ref{fig:profils} third column) are compatible with a dipole observed partially and also correspond to a dipole of a few G amplitude. We cannot however constrain the dipole tilt nor get a high confidence on the mean magnetic strength of the large-scale structure. The map given in Figure \ref{fig:maps} (third row) is indicative and fairly similar to the one obtained by \citet{gaulme10}.

\subsection{HD 130322}
Nine ESPaDOnS spectra of HD 130322 (K0V) were secured  in January 2008. The rotation period of this star is 26 days \citep{simpson10}, much longer than our observing run of 10 days. As a consequence, only one third of the stellar surface is observed. This makes difficult a full reconstruction of the magnetic topology, since we do not have observational constraints on the un-observed part of the star (see Appendix B in \citet{fares12}). The circular polarisation profiles are, however, significantly detected in all observing epochs. 

We adopted a value of 80$^\circ$ for the stellar inclination and reconstruct the map with a \chisqr~of 0.9 (Fig. \ref{fig:maps} fourth row). The circular polarisation profiles (Fig \ref{fig:profils}, fourth column) are well-fitted by a magnetic structure dominated by a dipole (only $\sim$16\% of the field energy is toroidal) of 2.5 G mean strength. Data over more than a full rotation period would be needed to confirm this result and could still reveal a more complex large-scale structure of the magnetic field.

\subsection{HD 189733}
Table \ref{obslog} includes three observational campaigns of HD 189733 (K2V) using ESPaDOnS and NARVAL in 2006, 2007 and 2008 for completion with respect to the target sample presented here. However, their analysis has already been published in \citet{moutou07} and \citet{fares10} and will not be repeated here. HD 189733 has a mainly toroidal surface magnetic field with a strength of 20 to 40 G. The stellar surface has a DR of $\dom=0.146\pm0.049$~\rpd. 
The field extrapolation up to the location of the planet has been derived by \citet{fares09} and \citet{cohen11}. The planet is found to cross different stellar field configurations along its orbit. This makes the reconnection events between stellar and planetary magnetic fields possible on fractions of the orbit. The planetary radio emission from magnetospheric interaction with the stellar wind varies along the orbit \citep{fares10}
 
\subsection{HD 179949}
Two epochs of ESPaDOnS observations of HD 179949 (F8V) have been discussed in \cite{fares12}.  The 2009 data set is part of a joined campaign with XMM and ground-based spectroscopic data taken simultaneously with the spectropolarimetric observations. The additional data are described in \citet{gaetano}. HD 179949 exhibits a weak and mainly poloidal magnetic field of a few G and a tilt of $\sim$ 70\degr. A DR of $\dom=0.216\pm0.061$~\rpd\ has been measured. In this case also, the field at the stellar surface has been extrapolated up to the planetary orbit, for further studies concerning modelling the interactions \citep{fares12}.

\subsection{Stars without detected fields}
\subsubsection{XO-3}

We have secured twenty independent observations of XO-3 (F5V) with ESPADONS in October 2009. Despite a high SNR for most spectra (12 out of 20 have SNR above $\sim$300), there was no detection of polarisation in the Stokes V profiles.

In order to quantify an upper limit for a magnetic field of XO-3, we propose the following analysis: \begin{enumerate}
\item We select a star with similar mass and a rotation period to XO-3, but for which we have a magnetic field detection. The reconstructed magnetic field of the chosen star is used as a magnetic topology model for X0-3. 
\item From this fake magnetic field, we calculate Stokes V profiles. We compare these profiles to the observed noise properties at the phases of our observations.
\item If the signal exceeds the noise and should have led to a detection, we decrease the field strength, without changing its topology. We repeat step 2 until the signal from the fake Stokes V profile is just about the noise level of the observations, at which a lower limit on the magnetic field of the star is derived.
\end{enumerate}

This analysis may give a reasonable order of magnitude of the maximum field strength excluded by our data for a chosen field topology. It must be noted, however, that other parameters may alter this value as the inclination (low impact), the temporal sampling of the observations and the field complexity. This attempt to quantify our non-detection should therefore not be over-interpreted.

In the case of XO-3 where we have numerous observations and a fast rotating star ($\vsini=18.3$~\kms), the detection limit has the most relevant significance. We injected signals corresponding to two stars where a magnetic topology has been deduced from previous observations, and relatively close in stellar properties to XO-3: HD 102195 (section \ref{102195}) and HD 179949 \citep{fares12}.  HD 179949 has an effective temperature close to that of XO-3 but has a Rossby number $> 1.0$, while HD 102195 has a lower mass than XO-3 but has, as XO-3, a Rossby number $< 1.0$ (magnetic fields show similar properties for stars in different Rossby regimes, see section \ref{discussion}).
A field with similar properties than HD 102195 (with a mean field strength $\sim$10 G) projected on the observation space of XO-3 remains undetected at 3-$\sigma$ except in one spectrum and represents a reasonable detection limit in the context of a HD~102195's 55\% toroidal magnetic field topology.
The field of HD 179949, as characterized from the 2009 ESPaDOnS campaign \citep{fares12} is undetectable in the signal of XO-3. We multiplied by 10 all components of this magnetic field and found that the fake Stokes V signatures would have been detected in 7 over the 15 best-quality spectra with a significance larger than 3-$\sigma$. Thus, a mostly (90\%) poloidal field with an amplitude of 20~G  would have been unambiguously detected.  We adopt this more conservative value for an upper limit for the magnetic field strength of XO-3 during the 2009 observation campaign.


\subsubsection{HAT-P-2 = HD 147506}
HAT-P-2 (F8) has been observed four times between 26 June and 1 July 2007. None of the spectra shows a detection of the magnetic field, with a mean rms noise level in the LSD profiles of $0.5\times10^{-4}$.
HAT-P-2 is a fast rotating star with an effective temperature of 6290K, so its properties closely match the ones of $\tau$~Boo. In order to investigate the detection limit of the magnetic field in our data, we thus used one of the magnetic configurations depicted for $\tau$~Boo, scaled the field strength and calculated the fake Stokes V signatures that would have been produced at our observing sampling. We find that a 75\% poloidal field of 40~G would have been detected in two over the four observed phases. The actual field is thus either of very different configuration, or of lower strength (or both).

\subsubsection{CoRoT-7}
Four spectra of CoRoT-7 (G9V) have been secured with NARVAL in January 2010. Due to the faint luminosity of the star (V magnitude=11.7), the field detection represents a real challenge, especially for NARVAL. The SNR of the profiles is ten times lower than for HAT-P-2. In all four spectra, there is a spurious detection in the Stokes V profile, also detected in the null-polarisation check profiles. Only the last exposure shows a marginal detection of the magnetic field, with a Stokes V profile slightly larger than the null profile. We applied the same strategy described for XO-3 and HAT-P-2, taking a dipole as the magnetic model for this star (similar to the topology of HD~46375). We compared the fake Stokes V signatures that a dipolar field would produce at our observing phases. We find that a dipole with more than 150G strength would have been significantly detected in two spectra over the four available ones, and beyond the marginal detection on 27 January 2008. The detection limit represents a poorer constraint than for XO-3 and HAT-P-2, because the star is of lower mass, a slower rotator, and the spectra have lower SNR.

\section{Discussion and summary}
\label{discussion}
A summary of the main characteristics of the stellar magnetic fields observed in this study is given in Table \ref{obslog}. When the field is not detected, the upper value derived as explained above is shown. The stars of this study, with stellar masses 0.8 to 1.4 \mSun, feature large-scale magnetic fields of 2 to 40 G. Except at two epochs for HD~189733 and HD~102195, all other  targets have mainly poloidal fields, with varying degrees of axisymmetry.\\

\begin{figure}
\hspace{-1.2cm}
\includegraphics[scale=0.4,angle=270]{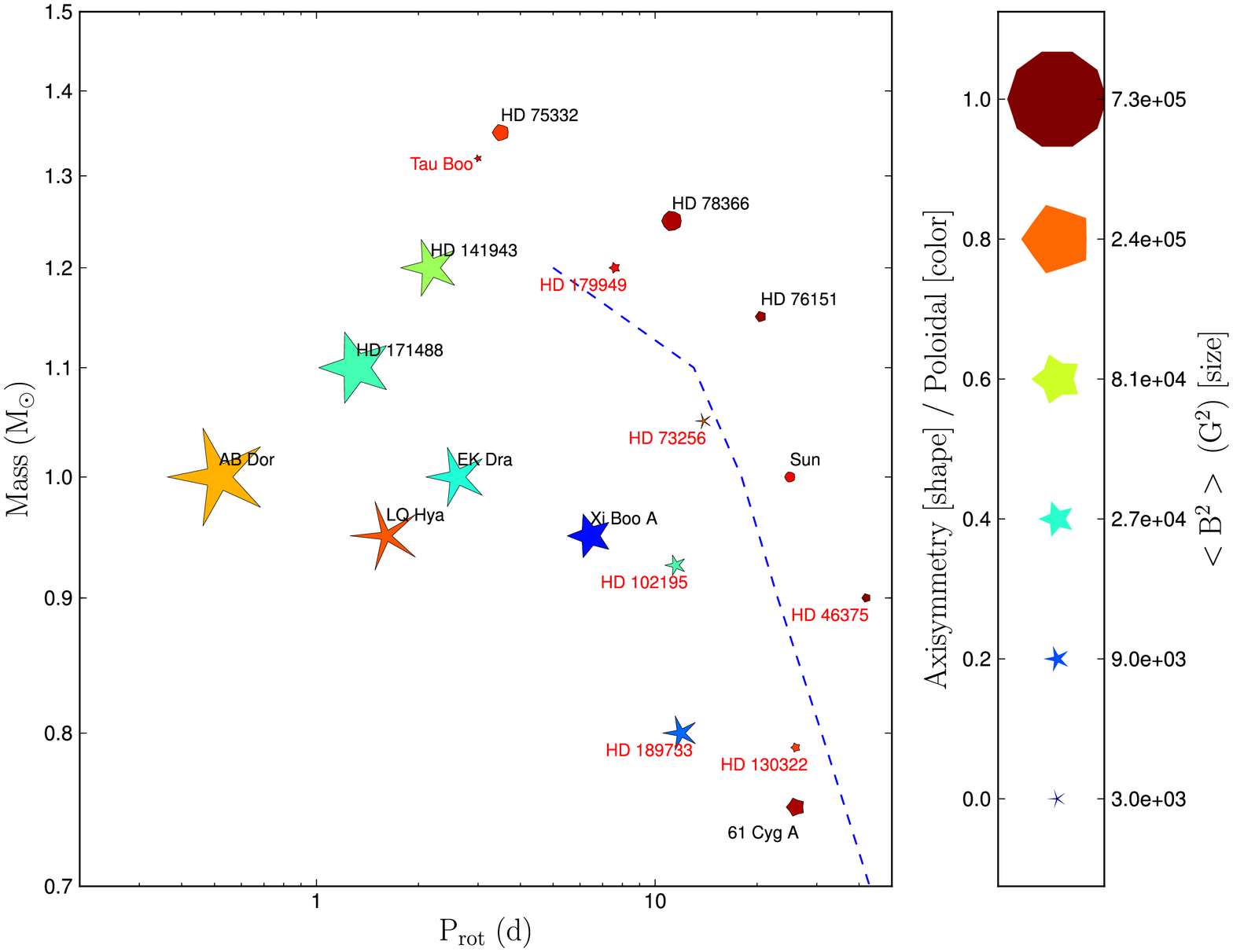}
\caption{Mass-rotation diagram of 18 reconstructed stellar magnetic fields (not including dM stars for instance). Planet host stars studied in this paper have their names indicated in red, while other stars without detected hot Jupiters have their names indicated in black (data from \protect\cite*{donati09}). The dashed line represent Rossby number = 1.0 (calculated using results of \protect\cite*{landin10}). The size of the symbol represents the field strength, its color the contribution of the poloidal component to the field, and its shape how axisymmetric the poloidal component is. For $\tau$~Boo, we show here the field for one epoch of observation  (mainly poloidal). Hot-Jupiter host stars do not seem to have different magnetic properties than the other stars.}
\label{fig:confuso}
\end{figure}

The presence of a giant planet at a small orbital distance is thought to have influences on the star. Empirical evidence suggest that tidal interactions can cause excess rotation of the parent star \citep{pont09}. \cite{cuntz00} suggest that these interactions can cause local instabilities in the tidal bulges and thus modify the local dynamo on action in these regions. \cite{cebron11a,cebron11b} present theoretical work on the effect of tidally-driven elliptical instability on star hosting a hot-Jupiter (HJ) and suggest that eventually these instabilities can produce a dynamo. In order to study possible peculiarities in the magnetic topologies of HJ host stars, one should compare their magnetic topologies to that of similar stars without detected closed-in giant planets. Figure \ref{fig:confuso} shows a mass-rotation plane including the magnetic properties of stars of our sample (whose names are shown in red), as well as other published stellar field properties (from \citealt{donati09}). A main transition appears in this plot for stars with masses above 0.5 \mSun\ : i) below Rossby number of $\sim$1, the large-scale field is mainly toroidal, with a non-axisymmetric poloidal component; ii) above Rossby number of $\sim$1, the field is weaker, poloidal and axisymmetric.  The long-term evolution of the magnetic fields should however be taken into account, with possible cycles, as for the Sun and a few other known examples \citep{donati08,fares09,Morgenthaler}. Our HJ host stars show similar field topologyies as the stars without a discovered close-in giant planet. The strength of their magnetic fields seems weaker, however, our sample is basically of stars chosen for radial velocity studies, they are less active than the other stars shown in Figure \ref{fig:confuso}. In addition, if the data quality is poor (poor S/N and poor phase coverage), the reconstructed  magnetic field strength is reduced below what would be reconstructed with higher quality data (see \citealt{fares12}). In order to comment of the field strength of planet-host stars, it is  necessary to enlarge our sample.

$\tau$~Boo's magnetic cycle is confirmed to be a short one. Observed between June 2006 and January 2011, the large-scale magnetic field of this star switches polarity yearly. However, the cycle duration might be of 2 years or of 8 months, as both periods are good solutions for the data we have. In the frequency domain, the 240 day period is the third harmonic of the 740 day period. The period of 740 days seems more likely. 
Previous studies of the chromospheric activity of this star found a period of 126 days persistant over 30 years \citep{maulik97,baliunas97}. If the relation between the activity cycle to the large-scale magnetic field cycle in stars is similar to that of the Sun, this favours the 8 month period for the magnetic cycle. However, such a relation is not known for stars (there is a lack of observed large-scale magnetic cycles), and thus it can not help us rule out one of the the two values we get. In order to favor one value over the other, we suggest dense observations of this star over a year, with at least 4 epochs of observations. Constraining the period will permit a comparison with the solar chromospheric/magnetic cycle behaviour. 

\cite{katja12} observed $\tau$~Boo in X-rays over six epochs (one observation in June 2003 and then 5 between June 2010 and June 2011). The star shows variability in X-rays, but a cyclic behaviour was not observed. They conjecture that the lack of X-ray cycles could be explained by either their sparse sampling, or that the polarity switch could be an artificial feature from the reconstruction method (ZDI) rather than being a real polarity switch. We note however that the lack of X-ray cycle does not rule out the presence of polarity switches (magnetic cycles), as have been predicted by theoretical works (\citealt{mcivor06b}, see also \citealt{baumann04,isik11}). Furthermore, in the particular case of $\tau$~Boo, Vidotto et al. (2012) simulated its stellar wind through the magnetic cycle, and studied mass-loss and angular-momentum loss rates, as well as X-ray emission measure and planetary radio emission. In their study, they used the magnetic maps from \citealt{fares09,donati08,catala07} as boundary condition for the stellar magnetic field (they thus considered the polarity switch in their model). They find that the emission measure does not vary during the cycle, suggesting that the quiescent X-ray emission of $\tau$~Boo does not change significantly over the cycle, agreeing with the findings of \cite{katja12}.

The goal of this work is to study for the first time the magnetic fields of a sample of planet-host stars. When possible, we reconstructed the stellar magnetic fields. We found a wide range of topologies. 
In order to check if these topologies are peculiar, we compared them to those of stars without detected hot-Jupiters. We found that these planet-host stars do not show peculiar magnetic behaviours. Our study shows that the stellar magnetic field topology is usually more complex than a simple dipole or quadrupole. Thus, it is essential to use the reconstructed maps in simulations (instead of modelling the field by a simple dipole). Our work is thus a basis for future star-planet interactions simulations, wind models, and simulations of radio and  X-ray emission. We will make our magnetic maps available to the public on the following link http://lamwws.oamp.fr/exo/starplanetinteractions/mtphs.

\section*{Acknowledgments}
RF acknowledges support from STFC consolidated grant ST/J001651/1. This work is based on observations obtained with ESPaDOnS at the Canada-France-Hawaii Telescope (CFHT) and with NARVAL at the T\'elescope Bernard Lyot (TBL). CFHT/ESPaDOnS are operated by the National Research Council of Canada, the Institut National des Sciences de l'Univers of the Centre National de la Recherche Scientifique (INSU/CNRS) of France, and the University of Hawaii, while TBL/NARVAL 
are operated by INSU/CNRS. We thank the CFHT and TBL staff for their help during the observations. We thank an anonymous referee for their comments.

\bibliography{paper_RFares}
\bibliographystyle{mn2e}

\appendix
\section{Journal of observations}
\label{appendix}
In this Appendix, the detailed journal of observations (Tables \ref{obsjournaltb}, \ref{obsjournal1} and \ref{obsjournal2}) is given for all data of the survey, except the ones already published in \citet{catala07, donati08,moutou07,fares09,fares10,fares12}. 

\begin{table*}
\vspace{1cm}
\caption{Journal of observations of $\tau$ Boo. Columns 1--12 sequentially list the star name, UT date, instrument used, the heliocentric Julian date (at mid-exposure),  the UT time (at mid-exposure), the complete exposure time, the peak signal to noise ratio (per 2.6~\kms\ velocity bin) of each observation (around 700~nm), the rotational cycle, the radial velocity (RV) associated with each exposure, the rms noise level (relative to the unpolarized continuum level $I_{\rm c}$ and per 1.8~\kms\ velocity bin) in the circular polarisation profile produced by Least-Squares Deconvolution (LSD), the longitudinal magnetic field and the false-alarm probability of the detection of the magnetic signature.}
\begin{tabular} {lccccccccccc} 
\hline
 Star          &       Date   &     Instrument&  HJD          &    UT   &   T$_{exp}$  &  S/N&  $\phi_{rot}$&   RV&  $\sigma_{LSD}$& B$_l$&      FAP\\
 		&		&			&(245 4000+)&(h:m:s)&(s)&			&	&(\kms) &(10$^{-4} I_c$)&(G)&\\
\hline
$\tau$ Boo & 27may09  & Narval & 979.39142 & 21:18:08 & 4$\times$700& 1030 & -2.531 & -16.163 & 0.33 & -0.0$\pm$1.5 &3.279$\times 10^{-01}$\\
$\tau$ Boo & 27may09  & Narval & 979.44073 & 22:29:08 & 4$\times$700& 1410 & -2.514 &-16.189 & 0.25 & -0.7$\pm$1.1 &5.612$\times 10^{-01}$\\
$\tau$ Boo & 27may09  & Narval & 979.55991 & 01:20:46 & 4$\times$600& 1430 & -2.475 & -16.276& 0.27 & 0.2$\pm$1.2 & 8.699$\times 10^{-01 }$\\
$\tau$ Boo & 28may09  & Narval & 980.40258 & 21:34:17 & 4$\times$700& 1650 & -2.194 &-16.821 & 0.22 & -0.4$\pm$1.0 &6.172$\times 10^{-01}$\\
$\tau$ Boo & 28may09  & Narval & 980.43825 & 22:25:39 & 4$\times$700& 1710 & -2.182 & -16.832& 0.22 & -0.6$\pm$1.0 &9.594$\times 10^{-01}$\\
$\tau$ Boo & 29may09  & Narval & 981.47684 & 23:21:18 & 4$\times$700& 1500 & -1.836 & -16.278& 0.26 & 1.0$\pm$1.2 &5.935$\times 10^{-07}$\\
$\tau$ Boo & 31may09  & Narval & 983.56341 & 01:26:07 & 4$\times$700& 1880 & -1.140 &-16.813 & 0.19 & 0.4$\pm$0.9 &8.525$\times 10^{-01}$\\
$\tau$ Boo & 01jun09  & Narval & 984.38396 & 21:07:46 & 4$\times$700& 1880 & -0.867 & -16.643& 0.19 & -1.8$\pm$0.8 & 2.259$\times 10^{-11 }$\\
$\tau$ Boo & 01jun09  & Narval & 984.41875 & 21:57:52 & 4$\times$700& 1830 & -0.855 &-16.620 & 0.19 & -0.8$\pm$0.9 & 5.689$\times 10^{-06 }$\\
$\tau$ Boo & 02jun09  & Narval & 985.36661 & 20:42:52 & 4$\times$700& 1800 & -0.539 & -15.948& 0.19 & 0.8$\pm$0.8 & 7.379$\times 10^{-02 }$\\
$\tau$ Boo & 02jun09  & Narval & 985.40138 & 21:32:56 & 4$\times$700& 1880 & -0.527 & -15.951& 0.19 & 1.6$\pm$0.8 & 1.343$\times 10^{-08 }$\\
$\tau$ Boo & 11jun09  & Narval & 994.40998 & 21:46:06 & 4$\times$700& 1380 & 2.475 & -16.520& 0.26 & 1.2$\pm$1.2 & 6.621$\times 10^{-01 }$\\
$\tau$ Boo & 11jun09  & Narval & 994.44475 & 22:36:11 & 4$\times$700& 1380 & 2.487 &-16.484  & 0.28 & 1.4$\pm$1.3 & 4.310$\times 10^{-01 }$\\
$\tau$ Boo & 12jun09  & Narval & 995.39500 & 21:24:37 & 4$\times$700& 1360 & 2.804 & -15.935& 0.30 & -1.7$\pm$1.3 & 2.871$\times 10^{-02 }$\\
$\tau$ Boo & 12jun09  & Narval & 995.42978 & 22:14:42 & 4$\times$700& 1480 & 2.815 &-15.923 & 0.28 & -1.9$\pm$1.2 & 8.053$\times 10^{-03 }$\\
$\tau$ Boo & 15jun09  & Narval & 998.54978 & 01:07:48 & 4$\times$600& 1690 & 3.855 &-15.940 & 0.24 & -2.3$\pm$1.1 & 8.522$\times 10^{-01 }$\\
\hline
$\tau$ Boo & 25jan10  & Narval & 1222.66429 & 03:55:40 & 4$\times$600& 1730 & -3.147 &-16.860 & 0.20 & -1.2$\pm$0.9 & 8.225$\times 10^{-07 }$\\
$\tau$ Boo & 25jan10  & Narval & 1222.73534 & 05:37:58 & 4$\times$600& 1790 & -3.126 &-16.837 & 0.19 & -2.3$\pm$0.9 &  8.424$\times 10^{-11}$\\
$\tau$ Boo & 25jan10  & Narval & 1222.76651 & 06:22:52 & 4$\times$600& 1590 & -3.116 &-16.825 & 0.22 & -0.6$\pm$1.0 & 2.225$\times 10^{-06        }$\\
$\tau$ Boo & 27jan10  & Narval & 1224.64686 & 03:30:20 & 4$\times$600& 1660 & -2.549 &-16.300 & 0.22 & 2.5$\pm$1.0 &   6.221$\times 10^{-05 }$\\
$\tau$ Boo & 27jan10  & Narval & 1224.71461 & 05:07:53 & 4$\times$600& 1490 & -2.528 &-16.361 & 0.24 & 2.9$\pm$1.1 &   3.485$\times 10^{-02 }$\\
$\tau$ Boo & 12feb10  & Narval & 1240.59580 & 02:14:58 & 4$\times$600& 1260 & 2.266 &-16.041 & 0.34 & 1.0$\pm$1.5 &    5.891$\times 10^{-01 }$\\
$\tau$ Boo & 13feb10  & Narval & 1241.63537 & 03:11:51 & 4$\times$600& 1020 & 2.580 &-16.747 & 0.43 & -1.0$\pm$1.9 &   6.981$\times 10^{-01}$\\
$\tau$ Boo & 13feb10  & Narval & 1241.71688 & 05:09:13 & 4$\times$600& 1130 & 2.605 &-16.813 & 0.37 & -2.2$\pm$1.7 &   4.960$\times 10^{-02}$\\
$\tau$ Boo & 14feb10  & Narval & 1242.60430 & 02:26:60 & 4$\times$600& 1570 & 2.872 &-16.833 & 0.23 & 1.2$\pm$1.0 &     7.482$\times 10^{-05}$\\
$\tau$ Boo & 14feb10  & Narval & 1242.71737 & 05:09:49 & 4$\times$600& 1630 & 2.907 &-16.737 & 0.21 & -0.4$\pm$0.9 &   5.608$\times 10^{-09}$\\
\hline
$\tau$ Boo & 14jan11  & Narval & 1576.70136 & 04:50:25 & 4$\times$600& 1730 & -2.267 &-16.956 & 0.19 & 1.1$\pm$0.9 &   9.118$\times 10^{-04}$\\
$\tau$ Boo & 15jan11  & Narval & 1577.69079 & 04:35:04 & 4$\times$600& 1490 & -1.968 &-16.357 & 0.23 & -0.4$\pm$1.0 &  1.731$\times 10^{-03 }$\\
$\tau$ Boo & 16jan11  & Narval & 1578.68566 & 04:27:34 & 4$\times$600& 1840 & -1.668 &-16.057 & 0.18 & -0.4$\pm$0.8 & 2.920$\times 10^{-01        }$\\
$\tau$ Boo & 22jan11  & Narval & 1584.67961 & 04:18:07 & 4$\times$600& 1670 & 0.142 &-16.147 & 0.20 & -1.4$\pm$0.9 & 5.847$\times 10^{-01 }$\\
$\tau$ Boo & 22jan11  & Narval & 1584.70997 & 05:01:50 & 4$\times$600& 1660 & 0.151 &-16.126 & 0.20 & -0.9$\pm$0.9 & 8.562$\times 10^{-02 }$\\
$\tau$ Boo & 22jan11  & Narval & 1584.74032 & 05:45:32 & 4$\times$600& 1530 & 0.160 &-16.117 & 0.22 & -1.7$\pm$1.0 &   6.900$\times 10^{-01}$\\
$\tau$ Boo & 23jan11  & Narval & 1585.68804 & 04:30:08 & 4$\times$600& 1760 & 0.446 &-16.390 & 0.18 & 0.8$\pm$0.8 &    3.481$\times 10^{-03}$\\
$\tau$ Boo & 23jan11  & Narval & 1585.72592 & 05:24:41 & 4$\times$600& 1860 & 0.458 &-16.428 & 0.18 & -0.3$\pm$0.8 &   2.009$\times 10^{-03}$\\
$\tau$ Boo & 23jan11  & Narval & 1585.75629 & 06:08:24 & 4$\times$600& 1810 & 0.467 &-16.450 & 0.18 & 1.7$\pm$0.8 & 2.857$\times 10^{-05  }$\\
$\tau$ Boo & 24jan11  & Narval & 1586.66487 & 03:56:39 & 4$\times$600& 1580 & 0.741 &-17.000 & 0.23 & -0.3$\pm$1.0 &   2.588$\times 10^{-03}$\\
$\tau$ Boo & 24jan11  & Narval & 1586.695230 & 04:40:22 & 4$\times$600& 1700 & 0.750 &-16.995 & 0.20 & -0.7$\pm$0.9 &   8.092$\times 10^{-03}$\\
$\tau$ Boo & 24jan11  & Narval & 1586.725580 & 05:24:05 & 4$\times$600& 1790 & 0.759 &-16.993 & 0.19 & 0.5$\pm$0.8 & 9.390$\times 10^{-03  }$\\
$\tau$ Boo & 25jan11  & Narval & 1587.666570 & 03:58:59 & 4$\times$600& 1780 & 1.043 &-16.391 & 0.18 & 0.3$\pm$0.8 & 3.947$\times 10^{-01  }$\\
$\tau$ Boo & 25jan11  & Narval & 1587.696930 & 04:42:42 & 4$\times$600& 1830 & 1.053 &-16.362 & 0.18 & -0.9$\pm$0.8 &   9.800$\times 10^{-02 }$\\
$\tau$ Boo & 25jan11  & Narval & 1587.727280 & 05:26:24 & 4$\times$600& 1820 & 1.062 &-16.336 & 0.18 & -0.4$\pm$0.8 & 8.442$\times 10^{-03 }$\\
$\tau$ Boo & 26jan11  & Narval & 1588.674650 & 04:10:30 & 4$\times$600& 930 & 1.348 &-16.197 & 0.37 & 2.8$\pm$1.6 &             5.712$\times 10^{-01 }$\\
$\tau$ Boo & 26jan11  & Narval & 1588.705000 & 04:54:12 & 4$\times$600& 1190 & 1.357 &-16.211 & 0.28 & 0.0$\pm$1.2 &    4.340$\times 10^{-01 }$\\
$\tau$ Boo & 26jan11  & Narval & 1588.735360 & 05:37:55 & 4$\times$600& 1650 & 1.366 &-16.227 & 0.20 & -0.4$\pm$0.9 &   5.440$\times 10^{-02}$\\
\hline
\end{tabular}
\label{obsjournaltb}
\end{table*}

\begin{table*}
\vspace{1cm}
\caption{Journal of observations of four stars for which the magnetic field was detected and analysed. }
\begin{tabular} {lccccccccccc} 
\hline
 Star          &       Date   &     Instrument&  HJD          &    UT time  &   T$_{exp}$  &  S/N&  $\phi_{rot}$&   RV&  $\sigma_{LSD}$& B$_l$&      FAP\\
 		&		&			&(245 4000+)&(h:m:s)&(s)&			&	&(\kms) &(10$^{-4} I_c$)&(G)&\\
\hline
HD 73256 & 19jan08  & ESPaDOnS& 484.89743 & 09:24:56 & 4$\times$560 & 410 & -0.252 & 30.322 & 0.78 & -2.6$\pm$1.5 & 8.171$\times10^{-3}$ \\
HD 73256 & 21jan08  & ESPaDOnS& 486.89575 & 09:22:25 & 4$\times$560 & 540 & -0.109 & 29.997 & 0.56 & 1.0$\pm$1.1 & 9.523$\times10^{-1}$ \\
HD 73256 & 22jan08  & ESPaDOnS& 487.91245 & 09:46:26 & 4$\times$560 & 600 & -0.036 & 30.313 & 0.51 & 5.3$\pm$1.0 & $<$$10^{-8}$  \\
HD 73256 & 23jan08  & ESPaDOnS& 488.95053 & 10:41:13 & 4$\times$700& 540 & 0.038 & 29.858 & 0.57 & 4.2$\pm$1.1 &$<$$10^{-8}$ \\
HD 73256 & 24jan08  & ESPaDOnS& 489.94860 & 10:38:24 & 4$\times$780 & 700 & 0.109 & 30.302 & 0.43 & 0.5$\pm$0.9 & 1.836$\times10^{-6}$\\
HD 73256 & 25jan08  & ESPaDOnS& 490.90682 & 09:38:12 & 4$\times$630 & 580 & 0.178 & 30.050 & 0.53 & -1.4$\pm$1.1 & 9.917$\times10^{-2}$  \\
HD 73256 & 27jan08  & ESPaDOnS& 492.91695 & 09:52:43 & 4$\times$900 & 580 & 0.321 & 30.356 & 0.47 & 0.3$\pm$0.9 &$<$$10^{-8}$ \\
HD 73256 & 28jan08  & ESPaDOnS& 493.91923 & 09:55:58 & 4$\times$900 & 390 & 0.393 & 29.877 & 0.69 & 4.6$\pm$1.4 & $<$$10^{-8}$  \\
HD 73256 & 30jan08  & ESPaDOnS& 495.92900 & 10:09:58 & 2$\times$900 & 360 & 0.536 & 30.131 & 0.83 & -0.7$\pm$1.6 &$<$$10^{-8}$ \\
\hline
HD 102195 & 19jan08  & ESPaDOnS& 484.95747 & 10:54:30 & 4$\times$540.0 & 270 & -0.228 & 2.209 & 1.31 & 5.4$\pm$2.4&$<$$10^{-8}$ \\
HD 102195 & 20jan08  & ESPaDOnS& 485.95843 & 10:55:46 & 4$\times$560.0 & 620 & -0.145 & 2.112 & 0.51 & 6.3$\pm$1.0&$<$$10^{-8}$ \\
HD 102195 & 21jan08  & ESPaDOnS& 486.95369 & 10:48:49 & 4$\times$560.0 & 560 & -0.062 & 2.132 & 0.57 & 5.5$\pm$1.2&$<$$10^{-8}$ \\
HD 102195 & 22jan08  & ESPaDOnS& 487.94096 & 10:30:22 & 4$\times$560.0 & 620 & 0.020 & 2.142 & 0.51 & 2.6$\pm$1.1&5.1 $\times10^{-5}$  \\
HD 102195 & 23jan08  & ESPaDOnS& 489.09831 & 14:16:50 & 4$\times$780.0 & 680 & 0.117 & 2.101 & 0.46 & 2.2$\pm$0.9&2.1 $\times10^{-1}$  \\
HD 102195 & 24jan08  & ESPaDOnS& 490.10308 & 14:23:34 & 4$\times$630.0 & 610 & 0.200 & 2.028 & 0.51 & 3.6$\pm$1.1&1.3 $\times10^{-5}$  \\
HD 102195 & 25jan08  & ESPaDOnS& 490.93814 & 10:25:58 & 4$\times$630.0 & 500 & 0.270 & 2.119 & 0.65 & 3.3$\pm$1.3&3.4 $\times10^{-2}$  \\
HD 102195 & 27jan08  & ESPaDOnS& 493.10187 & 14:21:30 & 4$\times$700.0 & 640 & 0.450 & 2.137 & 0.45 & -1.1$\pm$0.8&6.6$\times 10^{-3}$  \\
HD 102195 & 28jan08  & ESPaDOnS& 494.10213 & 14:21:46 & 4$\times$700.0 & 590 & 0.533 & 2.013 & 0.48 & 0.7$\pm$1.0&5.1 $\times10^{-6}$  \\
HD 102195 & 30jan08  & ESPaDOnS& 496.09029 & 14:04:31 & 4$\times$800.0 & 440 & 0.699 & 2.164 & 0.63 & 5.5$\pm$1.3& $<$$10^{-8}$\\
\hline
HD 46375 & 18jan08  & ESPaDOnS& 483.84291 & 08:06:50 & 4$\times$215 & 460 & 0.245 & -0.910 & 0.68 & 2.1$\pm$0.9 & $<$$10^{-8}$ \\
HD 46375 & 19jan08  & ESPaDOnS& 484.83247 & 07:51:51 & 4$\times$560& 550 & 0.269 & -0.912 & 0.56 & 2.4$\pm$0.8 &$<$$10^{-8}$ \\
HD 46375 & 21jan08  & ESPaDOnS& 486.83395 & 07:54:05 & 4$\times$560& 670 & 0.317 & -0.900 & 0.46 & 3.0$\pm$0.6 & $<$$10^{-8}$  \\
HD 46375 & 22jan08  & ESPaDOnS& 487.88189 & 09:03:10 & 4$\times$560& 670 & 0.342 & -0.953 & 0.46 & 3.1$\pm$0.6 & $<$$10^{-8}$ \\
HD 46375 & 23jan08  & ESPaDOnS& 488.91150 & 09:45:52 & 4$\times$780 & 770 & 0.366 & -0.992 & 0.39 & 2.5$\pm$0.5 & $<$$10^{-8}$ \\
HD 46375 & 24jan08  & ESPaDOnS& 489.90655 & 09:38:48 & 4$\times$780 & 780 & 0.390 & -0.920 & 0.39 & 3.0$\pm$0.5 & $<$$10^{-8}$ \\
HD 46375 & 25jan08  & ESPaDOnS& 490.87263 & 08:50:01 & 4$\times$630& 770 & 0.413 & -0.911 & 0.40 & 2.5$\pm$0.5 &$<$$10^{-8}$ \\
HD 46375 & 26jan08  & ESPaDOnS& 491.83784 & 07:59:59 & 4$\times$630& 650 & 0.436 & -0.956 & 0.43 & 2.7$\pm$0.6 &$<$$10^{-8}$ \\
HD 46375 & 27jan08  & ESPaDOnS& 492.79308 & 06:55:35 & 4$\times$800& 690 & 0.459 & -0.889 & 0.40 & 2.4$\pm$0.5 &$<$$10^{-8}$ \\
\hline
HD 130322 & 20jan08  & ESPaDOnS& 486.17067 & 16:07:04 & 4$\times$560.0 & 670 & -0.059 & -12.244 & 0.44 & 1.0 $\pm$1.1 & 4.482 $\times10^{-4}$ \\
HD 130322 & 21jan08  & ESPaDOnS& 487.17020 & 16:06:14 & 4$\times$560.0 & 620 & -0.020 & -12.292 & 0.48 & 2.3 $\pm$1.2 & 1.580$\times 10^{-2}$ \\
HD 130322 & 22jan08  & ESPaDOnS& 488.13040 & 15:08:47 & 4$\times$780.0 & 730 & 0.017 & -12.332 & 0.42 & 1.9 $\pm$1.0 & 1.219$\times 10^{-3}$ \\
HD 130322 & 23jan08  & ESPaDOnS& 489.13228 & 15:11:22 & 4$\times$700.0 & 700 & 0.055 & -12.342 & 0.43 & 3.2 $\pm$1.1 & 1.229 $\times10^{-2}$ \\
HD 130322 &  24jan08  & ESPaDOnS& 490.13394 & 15:13:37 & 4$\times$700.0 & 650 & 0.094 & -12.304 & 0.47 & 1.7 $\pm$1.2 & 1.129$\times 10^{-2}$ \\
HD 130322 &  25jan08  & ESPaDOnS& 491.11998 & 14:53:22 & 2$\times$630.0 & 410 & 0.131 & -12.227 & 0.68 & -0.8 $\pm$1.7 & 5.418 $\times10^{-1}$ \\
HD 130322 & 27jan08  & ESPaDOnS& 493.13465 & 15:14:12 & 4$\times$700.0 & 610 & 0.209 & -12.124 & 0.46 & 2.7 $\pm$1.1 & $<$$10^{-8}$  \\
HD 130322 &  28jan08  & ESPaDOnS& 494.13485 & 15:14:21 & 4$\times$700.0 & 540 & 0.247 & -12.124 & 0.51 & 3.2 $\pm$1.3 & 9.045$\times 10^{-6}$ \\
HD 130322 & 30jan08  & ESPaDOnS& 496.16789 & 16:01:39 & 2$\times$650.0 & 370 & 0.326 & -12.213 & 0.77 & 5.9 $\pm$1.9 & 2.070 $\times10^{-5}$ \\
\hline
\end{tabular}
\label{obsjournal1}
\end{table*}

\begin{table*}
\vspace{1cm}
\caption{Journal of observations of four stars for which the magnetic field is not detected.}
\begin{tabular} {lccccccccc} 
\hline
 Star          &       Date   &     Instrument&  HJD          &    UT time  &   T$_{exp}$  &  S/N&  $\phi_{rot}$&   RV&  $\sigma_{LSD}$ \\
 		&		&			&(245 4000+)&(h:m:s)&(s)&			&	&(\kms) &(10$^{-4} I_c$)\\
\hline
XO-3 & 14oct09  & ESPaDOnS& 753.914090 & 09:50:38 & 4$\times$860& 310 & 0.8911 & -12.663 & 0.90 \\
XO-3 & 14oct09  & ESPaDOnS& 754.007340 & 12:04:54 & 4$\times$860 & 250 & 0.9156 & -12.706 & 1.14 \\
XO-3 & 14oct09  & ESPaDOnS& 754.101130 & 14:19:58 & 4$\times$(860 & 310 & 0.9403 & -12.730 & 0.87 \\
XO-3 & 15oct09  & ESPaDOnS& 754.951580 & 10:44:35 & 4$\times$860 & 290 & 1.1641 & -11.998 & 0.93 \\
XO-3 & 15oct09  & ESPaDOnS& 755.030280 & 12:37:55 & 4$\times$860 & 310 & 1.1848 & -11.840 & 0.87 \\
XO-3 & 15oct09  & ESPaDOnS& 755.123800 & 14:52:35 & 4$\times$860 & 240 & 1.2094 & -11.590 & 1.14 \\
XO-3 & 16oct09  & ESPaDOnS& 755.929600 & 10:12:55 & 4$\times$860 & 130 & 1.4215 & -10.170 & 2.27 \\
XO-3 & 16oct09  & ESPaDOnS& 756.022450 & 12:26:37 & 4$\times$860 & 100 & 1.4459 & -10.398 & 3.38 \\
XO-3 & 16oct09  & ESPaDOnS& 756.129000 & 15:00:03 & 4$\times$860 & 50 & 1.4739 & -10.237 & 9.12 \\
XO-3 & 17oct09  & ESPaDOnS& 756.896140 & 09:24:42 & 4$\times$860 & 160 & 1.6758 & -12.451 & 1.88 \\
XO-3 & 17oct09  & ESPaDOnS& 757.009760 & 12:08:19 & 4$\times$860 & 90 & 1.7057 & -12.285 & 3.81 \\
XO-3 & 18oct09  & ESPaDOnS& 757.890810 & 09:17:01 & 4$\times$860 & 310 & 1.9376 & -12.478 & 0.89 \\
XO-3 & 18oct09  & ESPaDOnS& 757.999710 & 11:53:49 & 4$\times$860 & 290 & 1.9662 & -12.338 & 0.92 \\
XO-3 & 18oct09  & ESPaDOnS& 758.092700 & 14:07:44 & 4$\times$860 & 300 & 1.9907 & -12.175 & 0.91 \\
XO-3 & 19oct09  & ESPaDOnS& 758.924770 & 10:05:53 & 4$\times$860 & 300 & 2.2097 & -9.862 & 0.92 \\
XO-3 & 19oct09  & ESPaDOnS& 759.028770 & 12:35:39 & 4$\times$860 & 300 & 2.2370 & -10.000 & 0.89 \\
XO-3 & 19oct09  & ESPaDOnS& 759.122950 & 14:51:16 & 4$\times$860 & 290 & 2.2618 & -10.271 & 0.95 \\
XO-3 & 20oct09  & ESPaDOnS& 759.898910 & 09:28:38 & 4$\times$860 & 290 & 2.4660 & -12.298 & 0.94 \\
XO-3 & 20oct09  & ESPaDOnS& 759.987590 & 11:36:20 & 4$\times$860 & 290 & 2.4894 & -12.440 & 0.93 \\
XO-3 & 20oct09  & ESPaDOnS& 760.076200 & 13:43:56 & 4$\times$860 & 290 & 2.5127 & -12.540 & 0.94 \\
\hline
HAT-P-2 & 26jun07  & ESPaDOnS& 277.780050 & 06:38:01 & 4$\times$900& 470 & 0.6472 & -19.115 & 0.55 \\
HAT-P-2 & 27jun07  & ESPaDOnS& 278.797830 & 07:03:38 & 4$\times$800& 500 & 0.9082 & -19.254 & 0.51 \\
HAT-P-2 & 28jun07  & ESPaDOnS& 279.793500 & 06:57:25 & 4$\times$900 & 500 & 1.1635 & -19.490 & 0.51 \\
HAT-P-2 & 01jul07  & ESPaDOnS& 282.791670 & 06:54:50 & 4$\times$900 & 500 & 1.9322 & -19.164 & 0.53 \\
\hline
Corot-7& 05jan10  & Narval & 1202.515870 & 24:15:19 & 4$\times$2000& 120 & 52.9038 &31.264 & 3.08 \\
Corot-7& 06jan10  & Narval & 1203.495410 & 23:45:52 & 4$\times$2000& 80 & 53.2805 &31.226 & 4.87 \\
Corot-7 & 18jan10  & Narval & 1215.489870 & 23:38:10 & 4$\times$2000& 90 & 57.8938 &31.234 & 4.73 \\
Corot-7 & 27jan10  & Narval & 1224.459390 & 22:54:42 & 4$\times$2000& 130 & 61.3436 &31.188 & 2.87 \\
\hline
\end{tabular}
\label{obsjournal2}
\end{table*}

\end{document}